\documentclass[final,journal]{IEEEtran}
\IEEEoverridecommandlockouts
\usepackage{epsfig,amsmath,amssymb,amsfonts,bm}
\usepackage{cite}
\usepackage{color}
\usepackage{soul}
\usepackage{subfigure}
\usepackage{epstopdf}
\usepackage{overpic}
\graphicspath{{figs/}}
\usepackage{booktabs}
\usepackage{rotating}
\usepackage{stfloats}
\usepackage[colorlinks=false,
           linkcolor=red,
           anchorcolor=blue,
           citecolor=green
           ]{hyperref}
\allowdisplaybreaks[4]
\usepackage{cases}
\usepackage{rotating}
\usepackage{array}
\usepackage{multirow}
\usepackage{longtable}
\usepackage{booktabs}
\usepackage{newtxtext, newtxmath}
\usepackage{makecell}
\usepackage{tikz}
\usetikzlibrary{positioning, arrows.meta, shadows}
\usepackage{rotating}
\begin{document}

\title{Spatial-Spectral Chromatic Coding of Interference Signatures in SAR Imagery: Signal Modeling and Physical-Visual Interpretation}

\author{
Huizhang~Yang,
Chengzhi Chen,
Liyuan~Chen,
Zhongling~Huang,\\
Zhong~Liu,~\IEEEmembership{Member,~IEEE},
and 
Jian~Yang,~\IEEEmembership{Senior Member,~IEEE}
\thanks{
This work was supported in part by Natural Science Foundation of China under grant No. 62301259 and 62171224, the Natural Science Foundation of Jiangsu Province, China, under grant No.BK20221486, and the Fundamental Research Funds for the Central Universities under grant No.30924010914.

Huizhang Yang, Liyuan Chen, Chengzhi Chen, and Zhong Liu are with the School of Electronic and Optical Engineering, Nanjing University of Science and Technology, Nanjing 210094, China (e-mail: hzyang@njust.edu.cn).

Zhongling Huang is with the School of Automation, Northwestern Polytechnical University, Xian 710072, China (email: huangzhongling@nwpu.edu.cn).

Jian Yang is with the Department of Electronic Engineering, Tsinghua University, Beijing
100084, China.
}
}
\maketitle
\begin{abstract}

Synthetic Aperture Radar (SAR) images are conventionally visualized as grayscale amplitude representations, which often fail to explicitly reveal interference characteristics caused by external radio emitters and unfocused signals. This paper proposes a novel spatial-spectral chromatic coding method for visual analysis of interference patterns in single-look complex (SLC) SAR imagery. The method first generates a series of spatial-spectral images via spectral subband decomposition that preserve both spatial structures and spectral signatures. These images are subsequently chromatically coded into a color representation using RGB/HSV dual-space coding, using a set of specifically designed color palette. This method intrinsically encodes the spatial-spectral properties of interference into visually discernible patterns, enabling rapid visual interpretation without additional processing. To facilitate physical interpretation, mathematical models are established to theoretically analyze the  physical mechanisms of responses to various interference types. Experiments using real datasets demonstrate that the method effectively highlights interference regions and unfocused echo or signal responses (e.g., blurring, ambiguities, and moving target effects), providing analysts with a practical tool for visual interpretation, quality assessment, and data diagnosis in SAR imagery.
\end{abstract}
\begin{IEEEkeywords}
Synthetic Aperture Radar (SAR),  Interference Analysis, Interference Visualization, Spectral Analysis, Color Synthesis, Image Interpretation.
\end{IEEEkeywords}
\IEEEpeerreviewmaketitle

\section{Introduction}
Synthetic Aperture Radar (SAR) has emerged as an indispensable tool for remote sensing applications, ranging from environmental monitoring and disaster assessment to military reconnaissance and geophysical exploration. Its all-weather, day-and-night imaging capability provides unique advantages over optical systems, particularly in scenarios obscured by clouds, vegetation, or darkness. SAR systems capture complex-valued data, typically stored as single-look complex (SLC) images, which contain both amplitude and phase information\cite{cumming2005digital}. Conventionally, SLC images are visualized as grayscale amplitude representations, where pixel amplitude corresponds to the magnitude of the two-dimensional compressed echo signal\cite{cumming2005digital}. While this approach effectively highlights structural features like buildings, roads, and terrain textures, it inherently discards critical spectral characteristics embedded within the complex data.

A significant limitation of traditional grayscale visualization arises in the presence of radio interference \cite{tao2019mitigation} or other signal distortions (like azimuth ambiguities due to Doppler aliasing, point spread function blurring due to incorrect Doppler centroid estimation, and unfocused energy due to moving targets). Various interference signals manifest as anomalous patterns in SAR imagery\cite{yang2020mutual,yang2021bsf}, often corrupting data integrity and impeding accurate interpretation\cite{de2018spectrum}. In amplitude-based displays, interference may appear as bright streaks, saturated patches, or subtle artifacts that blend ambiguously with natural features. Crucially, grayscale representations fail to encode the \textit{spectral signature} of interference, which is intrinsically linked to its spatial-spectral properties. For instance, narrowband interference concentrates energy in specific frequency subbands, while wideband interference exhibits broader spectral dispersion\cite{tao2019mitigation}. Without explicit spectral chromatic coding, these distinctions remain invisible to the human observer, complicating efforts to diagnose, mitigate, or exploit interference phenomena.

Existing methods for interference visualization in SAR imagery primarily rely on time-frequency analysis or statistical anomaly detection\cite{zhang2011,yangMutualInterferenceSpaceborne,tao2015wideband,liSimultaneousScreeningDetection2022a,lvTwoStepApproachPulse2021}. Techniques such as short-time Fourier transform (STFT) can reveal localized interference patterns but often require manual parameter tuning and produce outputs that are challenging to integrate into standard image interpretation workflows. Polarimetric decomposition methods\cite{lee2017polarimetric}, though effective for distinguishing scattering mechanisms, are not designed to isolate frequency-specific interference signatures. Moreover, these approaches typically generate auxiliary data layers rather than directly enhancing the primary image display, thereby limiting their utility for rapid visual assessment. Consequently, there exists a critical gap in techniques that can \textit{naturally} embed interference characteristics into an intuitive, color-coded visualization framework without sacrificing spatial context or introducing computational complexity.

To address these limitations, this paper introduces a novel \textbf{spatial-spectral chromatic coding} method for SLC image visualization. The first core innovation lies in the generation of spatial-spectral image series of an SLC image via decomposing the range frequency spectrum into multiple spectral subbands, each corresponding to a distinct segment of the radar's bandwidth. By applying bandpass filters to the SLC or range-compressed data, we generate a set of spatial-spectral images that preserve both spatial structures and localized spectral energy distributions. The second core innovation is the chromatic coding of the spatial-spectral images in RGB and HSV space, using a series of carefully designed basis color palette under a “white” constraint on the clean region. The resulting visualization leverages the human visual system's sensitivity to color contrasts: interference-contaminated regions exhibit vivid hues due to non-uniform energy distribution across subbands, while clean areas---characterized by balanced spectral content---appear near-grayscale. This approach inherently encodes the interference's temporal and spectral properties into perceptually salient color patterns, enabling immediate visual discrimination between natural and anomalous signals.

The contributions of this work are summarized as follows:
\begin{enumerate}
    \item \textbf{Mathematical Modeling and Physical Mechanism Interpretation}: We establish mathematical models and reveal physical mechanisms to theoretically analyze chromatic responses to various interference types, including continuous-wave narrowband interference (CW-NBI), linear frequency modulation (LFM) signals, and unfocused signals.
    \item \textbf{Algorithmic Framework}: We present a mathematically rigorous yet computationally efficient pipeline for spatial-spectral images generation via spectrally equalized subband decomposition and chromatic coding in RGB/HSV dual-space using a set of reference color palette.
    \item \textbf{Sum-white Criterion for Color Palette Design}: We propose a sum-white constraint for designing the reference color palette for spatial-spectral chromatic coding, which effectively highlights interference regions as well as unfocused signals (unwanted artefacts like blurring, ambiguities, and moving targets responses) with high chromatic contrast, while keeps clean region achromatic.
    \item \textbf{Validation and Analysis}: Through experiments with real data including SIDS dataset\cite{SIDS}, we demonstrate that the method facilitates interference identification and analysis, providing a unified framework for visualizing, diagnosing, and interpreting diverse interference and unwanted artefacts in SAR imagery.
\end{enumerate}

The remainder of this paper is organized as follows. Section II formulates the research gaps by reviewing conventional SAR visualization, interference mitigation techniques, and color encoding strategies. Section III establishes mathematical models for interpreting chromatic responses to various interference types. Section IV details the spatial-spectral image generation method, including subband partitioning, spectral equalization, and multilooking speckle noise reduction. Section V presents the core spatial-spectral chromatic coding framework, covering RGB/HSV dual-space mapping, color palette design under the sum-white constraint, and brightness optimization.  Section VI validates the proposed method using real SAR datasets. Finally, Section VII concludes the paper and discusses future research directions.

\section{Problem Formulation}
The visualization and analysis of interference in SLC images have not been extensively studied. This section reviews three interconnected research domains: (1) standard SAR visualization techniques, (2) interference detection and mitigation methods, and (3) color-based encoding strategies for complex data. We highlight critical gaps that motivate the proposed spatial-spectral chromatic coding framework.

\subsection{Conventional SAR Visualization}
SLC images are traditionally visualized as grayscale amplitude representations, where pixel amplitude corresponds to the magnitude of complex-valued coherently compressed radar echoes. While intuitive for general scene interpretation, this approach obscures critical phase and spectral information. For instance, amplitude-only displays fail to distinguish between interference and natural scattering heterogeneity, leading to misinterpretation of artifacts as terrain features. Alternative techniques, such as phase-derived displays, preserve phase information but require expert knowledge for interpretation or remain impractical for large-scale analysis. Some pseudo-color mapping methods leverage multi-temporal or polarimetric data to enhance discrimination, yet these rely on auxiliary datasets unavailable in single-channel SLC acquisitions.

\subsection{Interference Detection and Mitigation}
Interference mitigation algorithms typically operate in the time, frequency, or time-frequency domain, exploiting interference's narrowband or impulsive characteristics. Notable methods include notch filtering in frequency or time-frequency domain\cite{Meyer2013,zhang2011},  eigenvalue decomposition\cite{ZhouEigensubspace2007}, subspace filtering\cite{yang2021bsf,yang2024robust}, spectral analysis based methods\cite{ren2018rfi,yang2021two,lv2023mitigate}, and sparse optimization\cite{Nguyen2015Radio,SuNarrow2017,huang2018narrowband,yang2020dictionary,lu2021accurate}. While effective for suppression, these approaches treat interference as noise to be removed rather than a feature to be analyzed. Consequently, they discard valuable diagnostic information about interference sources, such as carrier frequency, bandwidth, and temporal modulation. Machine learning techniques, including autoencoders and convolutional neural networks (CNNs), have shown promise in detecting interference patterns\cite{yuMulticlassRadioFrequency2018,fanInterferenceMitigationSynthetic2019,
artiemjewDeepLearningRFI2021a,
taoRadioFrequencyInterference2021,
liSemanticCognitionEnhancment2021a,
luAutomaticRFIIdentification2022,
taoRadioFrequencyInterference2022b,
cenSelfSupervisedLearningMethod2023,
zhaoIntelligentDetectionSegmentation2023a,
sorensenRADIOFREQUENCYINTERFERENCE2023a,s20102919} but require extensive labeled training data and lack transparency in decision-making. Crucially, none of these methods provide an intuitive visual representation of interference characteristics for human analysts.

\subsection{Color coding for Complex SAR Data}
Color-based visualization has been explored in SAR for multi-dimensional data fusion. Polarimetric SAR (PolSAR) decompositions (e.g., Pauli, Huynen, Cloude, Freeman-Durden) map scattering mechanisms to RGB channels, while interferometric SAR (InSAR) uses hue-saturation-value (HSV) models to encode phase and coherence. However, these techniques assume multi-channel or multi-temporal inputs, rendering them inapplicable to single-look complex images. 

\subsection{Research Gaps and Motivation}
Existing literature reveals three persistent limitations that impede comprehensive SAR interference analysis. First, conventional amplitude-based visualization discards phase and spectral signatures critical for interference characterization, resulting in significant information loss that obscures diagnostic features. Moreover, mitigation-focused algorithms typically erase interference artifacts entirely rather than preserving them for analytical examination, creating diagnostic opaqueness that hinders source identification. Compounding these issues, color encoding methods remain dependent on multi-channel data inputs, leaving single-channel SLC imagery underexploited despite its prevalence in operational scenarios.

To address these challenges, this work introduces a spatial-spectral chromatic coding framework that fundamentally reimagines interference visualization. The approach decomposes SLC range spectra into distinct subbands to isolate interference signatures while preserving their intrinsic spatial-spectral characteristics. By mapping subband energy distributions directly to RGB channels without requiring auxiliary data, the method generates intuitive color contrasts where interference manifests as chromatic deviations against a grayscale background. This transformation converts interference from an artifact to be removed into a visually interpretable feature, enabling rapid identification and analysis through human-perceptible patterns.

\section{Mathematical Model and Physical Mechanism for Chromatic Interpretation}
This section develops theoretical models  for characterizing the spatial-spectral properties of interference signals across the SAR processing chain. We analyze three representative interference types—continuous-wave narrowband interference, pulsed linear frequency modulated (LFM) interference, and unfocused signals—by tracing their evolution from raw data domain to focused single-look complex (SLC) image domain. The mathematical derivations provides theoretical foundations for interpreting the chromatic images produced by our algorithm presented in sections IV and V.
  \begin{figure*}
  \centering
  \subfigure[]{\includegraphics[width=4.5cm,height=5cm]{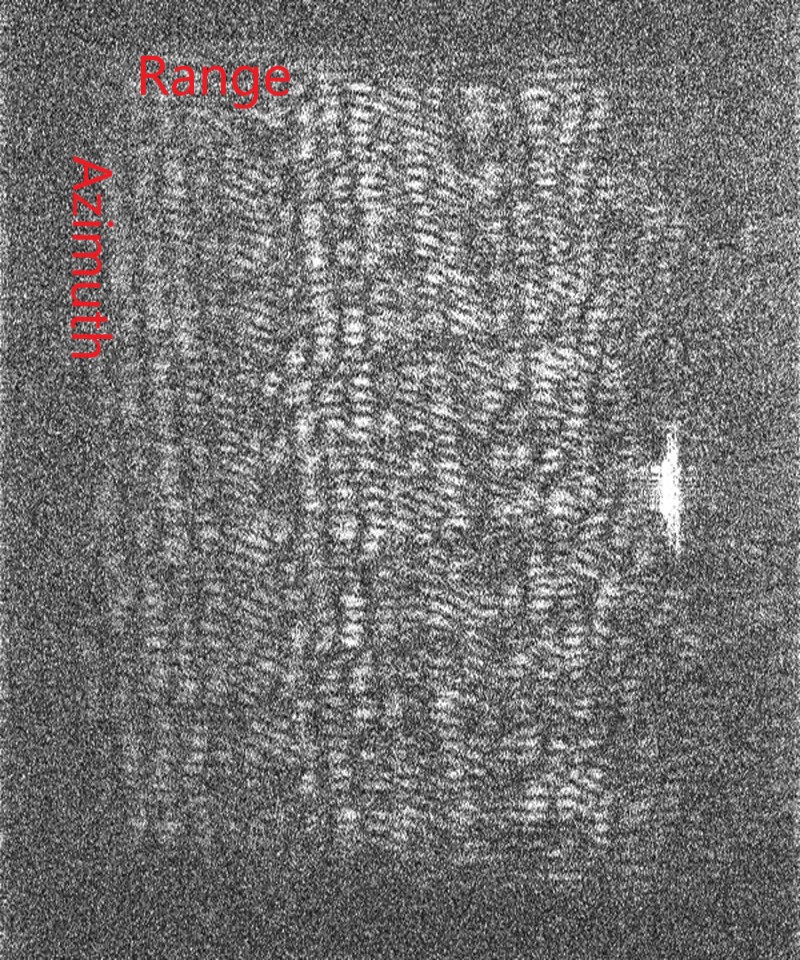}}\hspace{0.1cm}
    \subfigure[]{\includegraphics[width=4.5cm,height=5cm]{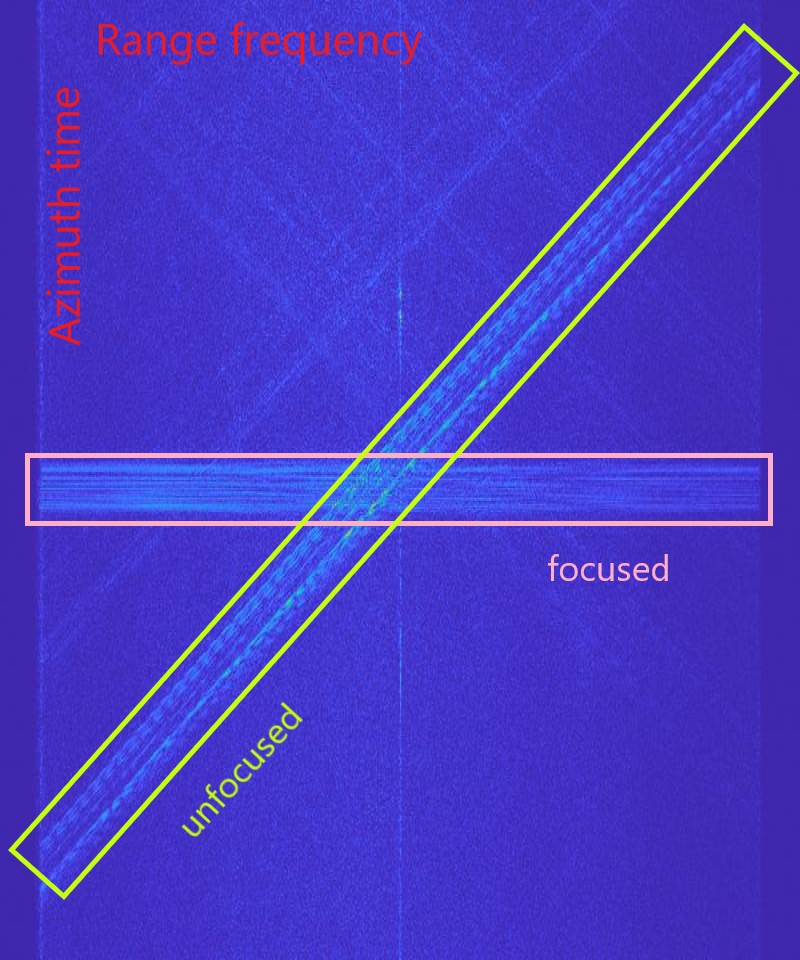}}\hspace{0.1cm}
      \subfigure[]{\includegraphics[width=4.5cm,height=5cm]{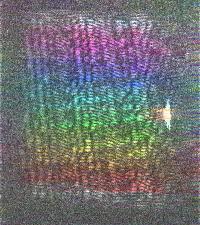}}
  \caption{The sloped spectrum of unfocused signal in the azimuth time - range frequency $(\eta, f_{\tau}^{\prime})$ domain. (a) Grayscale image, (b) the azimuth time - range frequency spectrum, (c) the chromatic coding results.}\label{fig:slop_spectrum}
\end{figure*}

\subsection{Continuous-wave Narrowband Interference}
Continuous-wave narrowband interference (CW-NBI) represents one of the most prevalent interference types in SAR systems, originating from sources such as communication signals, broadcast transmitters, or navigation systems. 

\subsubsection{Interference Signal Model} CW-NBI can be mathematically represented as a superposition of complex sinusoidal components in the raw data domain
\begin{equation}
y_{\text{CW-NBI}}(\eta, \tau) = \sum_{l=1}^{L} A_l(\eta) \exp\left[j2\pi f_l \tau + \phi_l(\eta)\right]
\end{equation}
where $\eta$ denotes azimuth slow time, $\tau$ is range fast time, $A_l(\eta)$ represents the time-varying amplitude of the $l$-th interference component, $f_l$ is its carrier frequency, and $\phi_l(\eta)$ is the phase term. The model captures the essential characteristics of CW-NBI: spectral concentration at specific frequencies $f_l$ and temporal continuity across the synthetic aperture.

\subsubsection{Omega-K Domain Transformation} The SAR focusing process, particularly the omega-k algorithm, establishes a critical link between raw data and SLC image domains. The omega-k method employs Stolt interpolation to transform raw data into the 2-D frequency domain, expressed as
\begin{equation} \label{eq:stolt}
\sqrt{(f_0+f_\tau)^2-\frac{c^2f_{\eta}^2}{4V_a^2}}=f_0+f_\tau^{\prime}
\end{equation}
where $f_0$ is the radar center frequency, $f_\tau$ is the range frequency variable, $f_\eta$ is the azimuth frequency (Doppler) variable, $V_a$ is the radar platform velocity, and $c$ is the speed of light. This equation describes the mapping from the raw data spectrum $(f_\eta, f_\tau)$ to the SLC image spectrum $(f_\eta, f_\tau^{\prime})$.
For typical SAR systems with low squint angles and narrow scene width, we derive an approximate stolt mapping
\begin{equation}
f_0+f_\tau^{\prime}\approx \frac{f_\tau}{\sqrt{1-\frac{c^2f_{\eta_c}^2}{4f_0^2V_a^2}}} + f_0 \sqrt{1-\frac{c^2f_{\eta}^2}{4f_0^2V_a^2}}
\end{equation}
where $f_{\eta_c}$ denotes the Doppler centroid frequency. This approximation reveals that the range spectral structure of CW-NBI remains largely invariant through the focusing process. Specifically, the spectral peaks at frequencies $f_{\tau}$ in the raw data domain are mapped directly to corresponding peaks in the SLC image domain, with its spatial extent determined by the SAR system parameters.

\subsubsection{Physical Mechanism of the Spatial-Spectral Chromatic Characteristic} The spectral preservation property causes CW-NBI to appear as distinct monochromatic artifacts in chromatically coded images. This characteristic enables effective identification in the chromatic feature space.

\subsection{Pulsed LFM Interference}
Pulsed linear frequency modulated (LFM) interference represents a critical challenge in modern SAR systems, particularly due to mutual interference between spaceborne SAR platforms. This section adapts the image-domain model from Yang et al. \cite{yang2021bsf} to characterize its spatial-spectral behavior.

\subsubsection{Interference Signal Model} Pulsed LFM interference originates from pulsed radar systems sharing the frequency band. In the raw data domain, a single LFM pulse can be expressed as
\begin{equation}\label{}
  p(t)={\rm rect}\left(\frac{t}{T_{\rm i}}\right)e^{j2\pi f_{\rm c} t+j\pi K_{\rm i} t^2},
\end{equation}
where ${\rm rect}(\cdot)$ is the rectangular function, $f_{\rm c}$ is the center frequency of the RFI subtracting the SAR center frequency, and $T_{\rm i}$, $K_{\rm i}$ are the pulse duration and FM rate of the RFI, respectively.

\subsubsection{Image-Domain Characteristics}  After SAR focusing, the LFM interference response in the SLC image domain is a two-dimensional LFM waveform characterized by\cite{yang2020mutual,yang2021two}
 \begin{equation}\label{lfm2d}
 \begin{split}
   \tilde{r}_l(\eta',t') = & a_l\delta(\eta-\eta_l) p(t-t_l)\ast h(\eta,t)\\
   = & a_l {\rm rect}\left(\frac{\eta'-\eta_l}{T_{\rm a}}\right)e^{-j\pi K_{\rm a} (\eta'-\eta_l)^2} \\
  & {\rm rect}\left(\frac{t'-t_l}{T_{\rm r}'}\right)e^{j2\pi f_{\rm c} (t'-t_l)+j\pi K_{\rm r}' (t'-t_l)^2},
   \end{split}
 \end{equation}
 where $K_{\rm r}'=\frac{K_{\rm r}K_{\rm i}}{K_{\rm i}-K_{\rm r}}$,  $T_{\rm r}'=T_{\rm i}|K_{\rm i}-K_{\rm r}|/K_{\rm r}$ and $K_{\rm r}$ is the chirp rate of SAR transmitted pulses. {Here we assume $K_{\rm i}\neq K_{\rm r}$}, which is the common case in spaceborne SAR imagery. This azimuth extent $T_{\rm a}$ states that a LFM interference pulse is smeared in the azimuth direction with a span of the synthetic aperture time.

\subsubsection{Physical Mechanism of the Spatial-Spectral Chromatic Characteristic} This spatial-spectral coupling causes pulsed LFM interference to exhibit continuous color transitions (e.g., red to purple gradients) along both the range direction in chromatically coded images. If the proposed chromatic coding method is applied to the azimuth direction, the LFM interference response will also exhibit continuous color transitions in the azimuth direction. The gradient direction and magnitude are directly related to the chirp rates $K_{\rm i}$ and $K_{\rm r}$ in range, and $K_a$ in azimuth. This behavior provides a distinctive signature for identifying pulsed LFM interference in the chromatic feature space, enabling differentiation from CW-NBI and other interference types.

\subsection{Unfocused Echoes}
Unfocused echoes arise from various sources including moving targets, Doppler centroid estimation errors, or Doppler aliasing due to smaller PRF than the Doppler bandwidth. Their distinct spatial-spectral characteristics significantly impact chromatic interpretation.

\subsubsection{Signal Model with Doppler Mismatch} In the presence of a Doppler shift $\Delta$ (due to inaccurate Doppler centroid estimation, Doppler aliasing, or target motion), the raw data spectrum of a point target deviates from the ideal case and is expressed as
\begin{equation}\label{eq:raw-spectrum}
X_{\text{raw}}(f_{\eta}, f_{\tau}) = \exp\left(j\frac{-4\pi R_0}{c}\sqrt{(f_0 + f_\tau)^2 - \frac{c^2(f_{\eta} + \Delta)^2}{4V_a^2}}\right)
\end{equation}
where $\Delta$ represents the Doppler shift relative to the stationary case. Here the constant amplitude, and LFM term $\frac{\pi f_{\tau}^{\prime}}{K_r}$ are removed by range matched filtering. Applying the Stolt interpolation (\ref{eq:stolt}) for SAR focusing, the echo data is mapped into the 2-d spectrum domain of the SLC image, which can be expressed as
\begin{equation}\label{eq:slc-spectrum}
X_{\text{slc}}(f_{\eta}, f_{\tau}^{\prime}) =  \exp\left(j\frac{-4\pi R_0}{c}\sqrt{\left(f_{0} + f_{\tau}^{\prime}\right)^{2} - \dfrac{c^{2}\Delta}{4V_{a}^{2}}\left(2f_{\eta} + \Delta\right)}\right)
\end{equation}
The existence of the Doppler shift $\Delta$ introduces additional range-azimuth coupling, and consequently its time-domain version are unfocused signal responses.

\subsubsection{2-D Spectral Domain Coupling Analysis} To analyze the coupling property, we perform a first-order Taylor expansion of the phase of Eq. (\ref{eq:slc-spectrum}) with respect to $f_{\eta}$
\begin{equation}\label{eq:taylor-expansion}
\begin{split}
\Phi_{\text{slc}}(f_{\eta}, f_{\tau}^{\prime}) \approx &\sqrt{\left(f_{0} + f_{\tau}^{\prime}\right)^{2} - \frac{c^{2}\Delta^2}{4V_{a}^{2}}} - \frac{c^{2}\Delta}{4V_{a}^{2}\sqrt{\left(f_{0} + f_{\tau}^{\prime}\right)^{2} - \frac{c^{2}\Delta^2}{4V_{a}^{2}}}}f_{\eta} \\
&\approx\sqrt{\left(f_{0} + f_{\tau}^{\prime}\right)^{2} - \frac{c^{2}\Delta^2}{4V_{a}^{2}}} - \frac{c^{2}\Delta\left(f_{0} + f_{\tau}^{\prime}\right)}{4V_{a}^{2}\sqrt{1 - \frac{c^{2}\Delta^2}{4f_0^2V_{a}^{2}}}}f_{\eta} 
\end{split}
\end{equation}
For focused echoes ($\Delta = 0$), the first-order terms vanish, yielding the ideal response. However, for unfocused signals ($\Delta \neq 0$), a significant first-order coupling term emerges:
\begin{equation}\label{eq:coupling-term}
 \widetilde{X}_{\text{slc}}(f_{\eta}, f_{\tau}^{\prime}) = \exp\left(j\frac{\pi R_0 c\Delta}{V_{a}^{2}\sqrt{1 - \frac{c^{2}\Delta^2}{4f_0^2V_{a}^{2}}}}f_{\eta}f_{\tau}^{\prime}\right)
\end{equation}

\subsubsection{Physical Mechanism of the Spatial-Spectral Chromatic Characteristic} The coupling term in Eq. (\ref{eq:coupling-term}) introduces a linear relationship between azimuth time and range frequency in the $(\eta, f_{\tau}^{\prime})$ domain, which exists as a sloped spectral line. This also occurs in the $(f_{\eta}, \tau)$ domain. This linear coupling causes azimuthal  smearing and continuous color transitions in chromatically coded images, where color gradient direction and magnitude are determined by $\Delta$:
\begin{itemize}
\item Positive $\Delta$ (e.g., approaching targets) produces blue-to-red gradients.
\item Negative $\Delta$ (e.g., receding targets) produces red-to-blue gradients.
\item Gradient steepness is proportional to $|\Delta|$.
\end{itemize}
It is notable that the color gradient direction of the unfocused signal is azimuthal, which is different from that of LFM interferences (range color gradient). Fig. \ref{fig:slop_spectrum} provides an example to illustrate the above phenomena of sloped spectral line and color transitions for unfocused signal in GaoFen-3 SLC data.

The above characteristic provides a critical signature for identifying unfocused signals in chromatically coded images. Unlike CW-NBI's monochromatic appearance or pulsed LFM's range continuous gradients, unfocused signals exhibit distinct azimuthal color transitions that correlate with the underlying Doppler shift. This property enables their differentiation from stationary scene content and other interference types.

\section{Part 1 of the Algorithm: Spatial-Spectral Images Generation via Fourier Analysis}

In the interference visualization method proposed in this work, spatial-spectral image generation serves as a critical preprocessing step for subsequent chromatic coding. The image generation is implemented by means of subband partition, and forward/inverse Fourier transforms. By dividing the single-look complex (SLC) SAR image into multiple subbands along the range frequency domain, localized spectral features of interference signals can be effectively extracted. These features subsequently provide multi-channel inputs for chromatic coding. This section systematically elaborates the mathematical principles of subband partitioning, spectral equalization procedures, and multilooking noise reduction strategies, which forms the part 1 of our algorithm.

\subsection{Mathematical Principles of Range-Frequency Subband Partitioning}

The frequency domain characteristics of SAR signals along the range direction determine the energy distribution of interference components. For a given SLC image, its frequency-domain representation can be expressed as $S(t_{\rm a}, f) $ where $ f $ denotes the range frequency variable and $ t_{\rm a} $ represents the azimuth coordinate. Assuming a sampling rate $ f_s $ and baseband bandwidth $ B $, the frequency domain spans $ [-B/2, B/2] $. To extract localized spectral features, the domain is partitioned into $ N_{\text{sub}} $ equal-width subbands. The bandwidth of each subband is calculated as:
\begin{equation}
\Delta B = \frac{B}{N_{\text{sub}}}.
\label{eq:subband_width}
\end{equation}
The frequency range for the $ k $-th subband is defined by:
\begin{equation}
f_k \in \left[-\frac{B}{2} + (k-1)\frac{B}{N_{\text{sub}}}, -\frac{B}{2} + k\frac{B}{N_{\text{sub}}}\right].
\label{eq:freq_range}
\end{equation}
This partitioning mechanism maps spectral energy distributions to multiple channels, thereby providing spatial-spectral features for subsequent color mapping. 

\begin{figure}
  \centering
  \includegraphics[width=7cm]{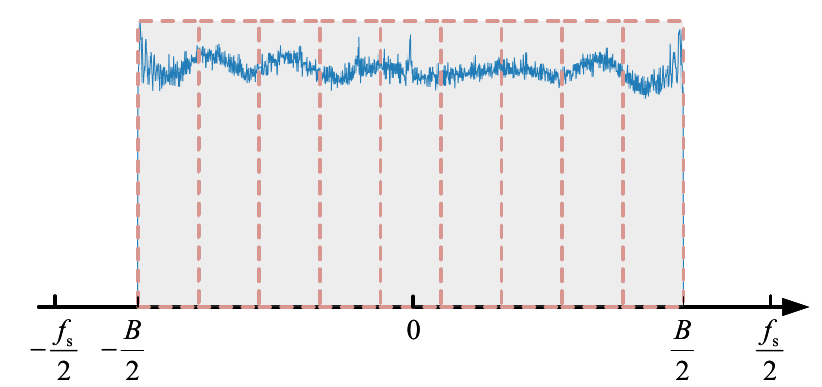}
  \caption{Illustration of spectral subband segmentation.}\label{fig:subband_demo}
\end{figure}

\subsection{Spectral Equalization: Compensating Windowing-Induced Distortions}

\begin{figure}
  \centering
{\includegraphics[width=6cm]{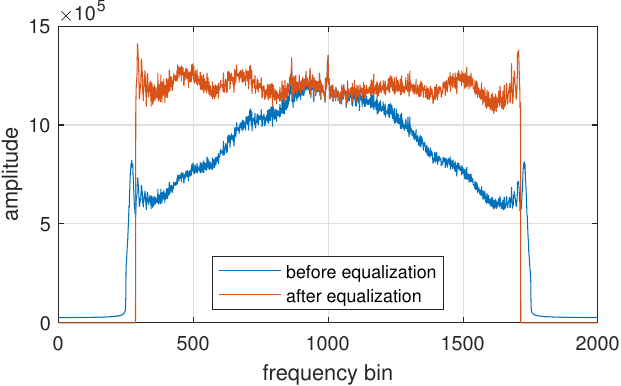}}

  \caption{Illustration of spectrum equalization. }\label{fig:spec_eq}
\end{figure}

Windowing operations (e.g., Hamming windows) in practical SAR imaging introduce spectral distortions characterized by amplitude attenuation at band edges. To address this, we propose a cosine window compensation method mathematically expressed as:
\begin{equation}
x(n) = a - (1 - a)\cos\left(2\pi \frac{n}{L_{\text{valid}}}\right),
\label{eq:cos_win}
\end{equation}
where $ a \in [0.5, 1] $ denotes the window coefficient and $ L_{\text{valid}} $ represents the effective spectral length. This method compensates for windowing-induced attenuation in the subband spectra by the window function. Experimental implementation demonstrates that when $ a = 0.75$ and the oversampling rate $ \xi = 1.2 \sim 1.5 $, the spectrum of Sentinel-1 data can be well equalized to be flat.  Fig. \ref{fig:spec_eq} shows an example of spectrum equalization processing.

\subsection{Multilooking Noise Reduction: Mitigating Speckle Effects}

Speckle noise in spatial-spectral images cause intensity fluctuation across subbands, which, if not suppressed, will significantly degrade the visualization quality of both interference and clean regions. To address this, we implement speckle reduction via multilooking processing through local averaging in both azimuth and range directions. With multilooking parameters $ \text{look} = [n,m] $, where $ m $ and $ n $ are odd integers denoting the number of looks in azimuth and range respectively, the mathematical formulation is:
\begin{equation}
I_{\text{ml},k}[i,j] = \frac{1}{n\,m} \sum_{p=-(n-1)/2}^{(n-1)/2} \sum_{q=-(m-1)/2}^{(m-1)/2} I_{k}(i+p, j+q),
\label{eq:multilook}
\end{equation}
where $ I_k $ represents the $ k $-th spatial-spectral image. The intensity of speckle pixels after multilook processing follows a Gamma distribution, expressed as:
\begin{equation}\label{eq:gamma_distribution}
  f(I;\sigma,\tilde{L})=\frac{1}{\Gamma(\tilde{L})}\left(\frac{\tilde{L}}{\sigma}\right)^{\tilde{L}}  
  I^{\tilde{L}-1}\exp\left(-\frac{\tilde{L}\, I}{\sigma}\right)
\end{equation}
where $\tilde{L}$ denotes the equivalent number of looks (ENL), and $\sigma$ represents the variance of the complex-valued pixels before multilook processing. When adjacent pixels are independent and identically distributed complex Gaussian variables, $\tilde{L}$ equals the actual number of averaged pixels $L=n\cdot m$. In practice, oversampling (e.g., 1.2 ) introduces spatial correlation between neighboring pixels, violating the independence assumption. Additionally, spatial variations in terrain scattering coefficients invalidate the identical distribution hypothesis. These factors result in $\tilde{L} < L$. The equivalent number of looks $\tilde{L}$ can be estimated using the method of moments. Experimental results demonstrate that a proper setting like $ \text{look} = [9, 9] $ achieve good results for subsequent chromatic coding processing. 

In summary, the spatial-spectral image generation workflow comprises five sequential stages:
\begin{enumerate}
\item \textbf{Spectral Equalization}: Apply cosine window compensation to each subband spectrum, yielding equalized spectra $ S_k'(t_{\rm a}, f) $.
    \item \textbf{Frequency Domain Partitioning}: Perform fast Fourier transform (FFT) on the SLC image to obtain frequency-domain data $ S(t_{\rm a}, f) $.
\item \textbf{Subband Extraction}: Divide the frequency domain into $ N_{\text{sub}} $ subbands according to $ N_{\text{sub}} $, extracting individual subband spectra $ S_k(t_{\rm a}, f) $.
\item \textbf{Inverse Transformation}: Conduct inverse FFT on the equalized spectra to generate spatial-spectral images $ I_k(t_{\rm a}, t) $.
\item \textbf{Multilooking Processing}: Apply multilooking to $ I_k(t_{\rm a}, t) $ to produce speckle-reduced spatial-spectral images $ I_{\text{ml},k}(t_{\rm a}, t) $.
\end{enumerate}
For $ N_{\text{sub}} = 9 $, this process generates nine spatial-spectral images covering localized spectral regions. These images serve as multi-channel inputs for subsequent chromatic coding through weighted superposition.

\section{Part 2 of the Algorithm: Spatial-Spectral Chromatic Coding in RGB and HSV Spaces}

The transformation of SAR subband information into a visually interpretable color representation constitutes a pivotal component of the proposed interference visualization framework. This section elaborates on the design of the RGB color mapping model, detailing the theoretical foundation, implementation workflow, and physical interpretation of the resulting color imagery. By systematically converting multi-channel subband data into a single composite color image, this method enables intuitive discrimination between interference-affected and clean regions through distinct chromatic signatures.
    
\subsection{Design of the Color Mapping Model}
The fundamental objective of the proposed color mapping strategy is to translate the sequence of spatial-spectral images, generated through range-frequency partitioning of the SLC data, into a composite color image that preserves the spectral characteristics of interference signals. This transformation relies on a carefully designed color space model, precise normalization procedures, and a robust synthesis mechanism to ensure both visual clarity and physical consistency.

\subsubsection{The Sum-White Constraint on Reference Color Palette Design}
We define a set of $N_{\text{sub}}$ distinct basis colors, ${\mathbf{c}_0, \mathbf{c}_1, \dots, \mathbf{c}_{N_{\text{sub}}-1}}$, where each color $\mathbf{c}_i$ is a vector in the RGB space, $\mathbf{c}_i = (r_i, g_i, b_i)$. The selection of these basis colors is critical for ensuring that different spectral components are visually distinguishable. A practical and effective approach is to sample the colors uniformly along the hue circle in the HSV (Hue, Saturation, Value) color space, ensuring maximum perceptual separation. For the method to be effective, the basis colors are constrained such that their summation approximates white, which can be expressed as:
\begin{equation}
\sum_{i=k}^{N_{\text{sub}}} \mathbf{c}_k \approx (r, r, r) 
\label{eq:sum-white}
\end{equation}
where $r$ is a constant number in $[0,N_{\text{sub}}]$. This constraint ensures that a pixel with a perfectly flat spectrum (i.e., equal intensity across all sub-bands, $I_1 = I_2 = \dots = I_{N_{\text{sub}}}$) will be rendered as a shade of gray, corresponding to an absence of spectral features. The physical significance of this constraint lies in its ability to establish a critical baseline: when spectral energy is uniformly distributed across subbands, the resulting achromatic representation provides a visual reference against which interference-induced chromatic variations can be effectively contrasted.

\subsubsection{HSV Color Space and Reference Color Palette}
\begin{figure}
\centering
{\includegraphics[width=4cm]{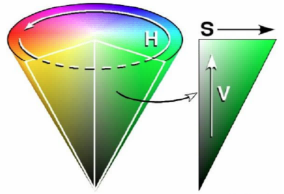}}
\caption{The HSV color model.}\label{fig:hsv_model}
\end{figure}

To achieve intuitive and physically meaningful color representation, the HSV (Hue-Saturation-Value) color model (as shown in Fig. \ref{fig:hsv_model}) is adopted as the foundation for constructing the reference color palette. The HSV model offers a natural alignment between perceptual color attributes and signal properties: the \emph{hue} parameter corresponds directly to the spectral frequency distribution, allowing different frequency bands to be mapped to distinct colors such as red ($H=0$), green ($H=0.33$), and blue ($H=0.67$); the \emph{saturation} reflects the degree of energy concentration across subbands, with interference regions exhibiting high saturation due to their non-uniform spectral distribution; and the \emph{value} component represents the normalized intensity, capturing the signal strength of both interference and echo in the final image.

\begin{figure}
\centering
\subfigure[]{\includegraphics[width=7cm]{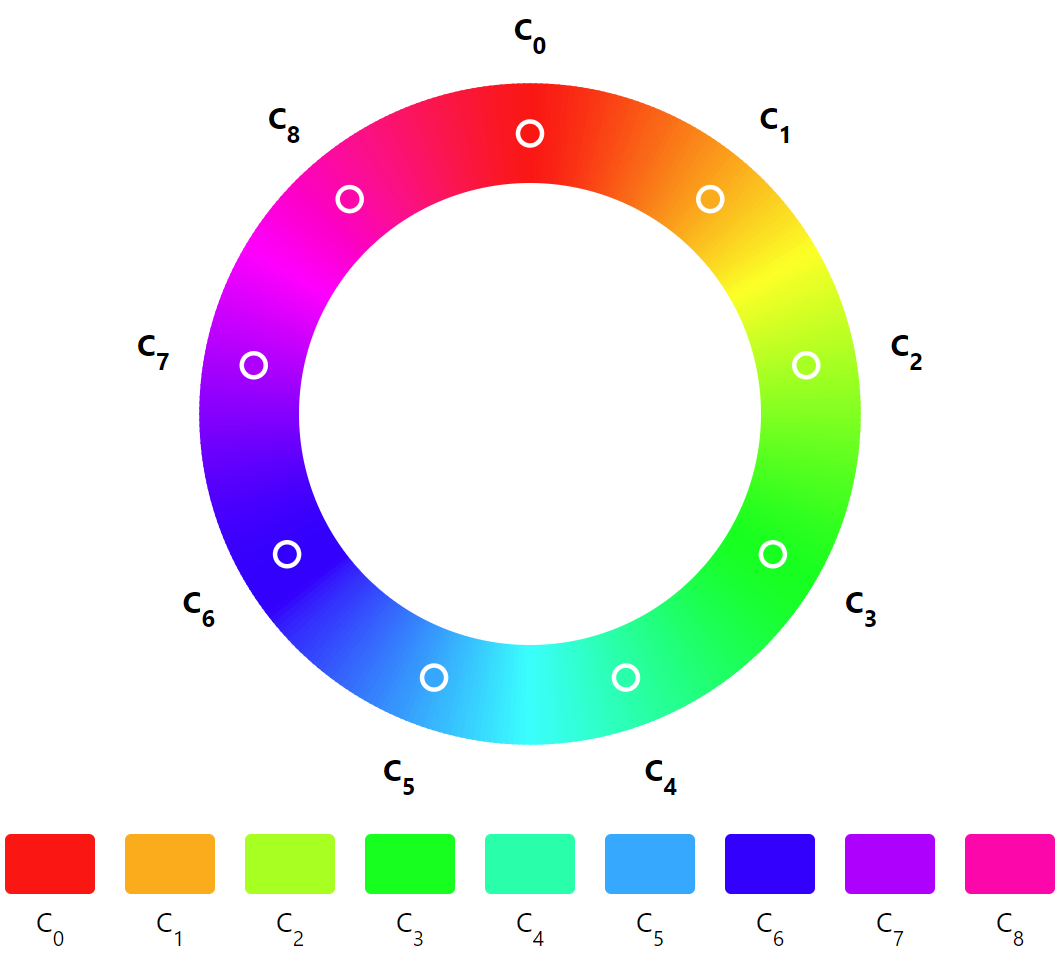}}
\caption{Reference color palettes obtained by uniformly sampling the Hue value in $[0,1]$. }\label{fig:color_pal}
\end{figure}

The reference color palette is constructed under specific constraints to maximize color contrast and maintain physical interpretability. First, the saturation of all reference colors is set to unity ($S_k = 1$), which amplifies chromatic differences between subbands and enhances the visibility of interference signatures. Second, the hue values are distributed uniformly across the color spectrum according to the number of subbands $N_{\text{sub}}$, following the relationship:
\begin{equation}
H_k = \frac{k}{N_{\text{sub}}}, \quad k = 0, 1, \dots, N_{\text{sub}}-1.
\label{eq:hue_distribution}
\end{equation}
For instance, with $N_{\text{sub}} = 9$, the hue progresses smoothly from red ($H=0$) through the spectrum to a deep blue ($H=0.889$), creating a perceptually uniform color gradient. Finally, the value component is normalized to unity ($V_k = 1$) for all colors, preventing any unintended brightness bias in the composite image. The reference color for the $k$-th subband is thus defined as:
\begin{equation}
\mathbf{C}_k = \left(H_k, S_k, V_k\right) = \left( \frac{k}{N_{\text{sub}}}, 1, 1 \right).
\label{eq:reference_color}
\end{equation}
The obtained set of reference color palette are visualized in Fig. \ref{fig:color_pal}. Crucially, this design inherently satisfies the sum-white constraint introduced earlier, as the uniform hue distribution combined with unit saturation and value ensures that the vector sum of the resulting RGB colors approximates $r(1,1,1)$. The above HSV representation is subsequently converted to the RGB color space using standard transformation algorithms (e.g., \texttt{hsv2rgb.m}), i.e., 
\begin{equation}
\mathbf{c}_k = \text{hsv2rgb}(\mathbf{C}_k), 
\label{eq:C-c}
\end{equation}
generating a set of $N_{\text{sub}}$ distinct reference colors that serve as the basis for the final image synthesis.

\subsubsection{Subband Intensity Normalization}

Prior to color synthesis, the intensity values of individual spatial-spectral images must be normalized to ensure balanced contribution across all frequency channels. This process begins with computing the amplitude of each multilooked spatial-spectral image $I_{\text{ml},k}[i,j]$, yielding $A_k[i,j] = |I_{\text{ml},k}[i,j]|$. A subband-domain normalization is then applied, scaling all subband amplitudes relative to the maximum observed value across the entire subband sequence:
\begin{equation}
\tilde{A}_k[i,j] = \frac{A_k[i,j]}{\max_{k'} A_{k'}[i,j]}.
\label{eq:global_normalization}
\end{equation}
This operation ensures that the strongest subband has a normalized amplitude of unity, with all other subbands scaled proportionally.

\subsubsection{Chromatic Coding in RGB Space}

The spatial-spectral features are chromatically coded in our images  through a weighted superposition of the normalized subband amplitudes and their corresponding reference colors. For each subband $k$, the normalized amplitude $\tilde{A}_k[i,j]$ is multiplied by the RGB components of its reference color $C_k$, producing a weighted color channel:
\begin{equation}
I_{\text{RGB},k}[i,j] = \big(R_k[i,j],G_k[i,j],B_k[i,j]\big) = \tilde{A}_k[i,j] \cdot \mathbf{c}_k.
\label{eq:weighted_channel}
\end{equation}
The resulting image $I_{\text{RGB},k}[i,j]$ contains three channels with values 
\begin{align}\label{eq:sumRGB}
&R_k[i,j]=\tilde{A}_k [i,j] r_k\\
&G_k[i,j]=\tilde{A}_k [i,j]g_k\\
&B_k[i,j]=\tilde{A}_k [i,j]b_k
\end{align}
 The composite RGB image is then formed by summing the contributions from all subbands:
\begin{equation}
I_{\text{RGB}}[i,j] = \big(R[i,j],G[i,j],B[i,j]\big)= \frac{1}{N_{\text{sub}}} \sum_{k=0}^{N_{\text{sub}}-1} I_{\text{RGB},k}[i,j].
\label{eq:composite_image}
\end{equation}
The division by $N_{\text{sub}}$ ensures that the total energy contribution remains normalized, preventing color saturation and maintaining visual balance across the image.

%
%
%
%
%

\subsubsection{Brightness Coding in HSV Space} In Eq. (\ref{eq:global_normalization}), the amplitude normalization operation only captures the spectral energy distribution, which does not contain the spatial brightness information.  To fix this, we perform brightness coding in the HSV space. Specifically, the obtained RGB image is then converted to HSV space, i.e.,
\begin{equation}
I_{\text{HSV}}[i,j] = \big(H[i,j],S[i,j],V[i,j]\big) = \text{rgb2hsv}(I_{\text{RGB}}[i,j]), 
\label{eq:I-rgb-hsv}
\end{equation}
 where the value channel $V[i,j]$, i.e., pixel brightness, is replaced by the multilooked SAR amplitude $|\tilde{I}_{\text{ml}}[i,j]|$. Referring to the HSV color model in Fig. \ref{fig:hsv_model}, coding the brightness in the HSV's value channel is consistent with traditional grayscale visualization methods, where week targets are dark and strong targets are bright. In SAR imagery, pixel amplitudes typically exhibit an extensive dynamic range, with the majority concentrated in the lower end of the amplitude histogram and only a small fraction extending into the higher values. Therefore, we  clip the amplitude at the $p$-th percentile for brightness coding:
\begin{equation}
V[i,j] = \min\left( \frac{|\tilde{I}_{\text{ml}}[i,j]|}{\text{percentile}(|\tilde{I}_{\text{ml}}|, p)}, 1 \right)
\label{eq:value_channel}
\end{equation}
This operation leads to improved images's dynamic range, and ensures the visibility of weak scatterers (prevents them from being too dark). The HSV image is then converted back to RGB space for final output representation.

\subsection{Physical Mechanism of Interference Chromatization}

The effectiveness of the proposed color mapping lies in its ability to differentiate interference-affected regions from clean areas through distinct chromatic patterns. This differentiation arises from fundamental differences in the spectral energy distribution between these two types of regions.

\subsubsection{Chromatic Characteristics of Interference Regions}

Interference signals, whether narrowband or wideband, typically exhibit a highly non-uniform energy distribution across the subbands. This concentration of energy in specific frequency bands results in significantly higher normalized amplitudes $\tilde{A}_k[i,j]$ for those subbands compared to the others. Due to the one-to-one correspondence between subband frequency and hue $H_k$, the dominant subband imparts its characteristic color to the interference region. For example, if the energy is concentrated in the third subband for $N_{\text{sub}} = 9$ (associated with green), the affected region will appear predominantly green in the composite image. When the interference spans multiple subbands, the resulting color is a mixture determined by the relative strengths of the contributing subbands, potentially yielding hues such as cyan or yellow. This chromatic signature provides a direct visual indication of the interference's spectral characteristics.

\subsubsection{Achromatic Characteristics of Clean Regions}

In contrast, the energy distribution of clean SAR signals is approximately uniform across all subbands. Consequently, the normalized amplitudes $\tilde{A}_k[i,j]$ are nearly constant (denoted as $a[i,j]$ for subsequent analysis) for all $k$. When these uniform amplitudes are combined with the uniformly distributed hues of the reference color palette, the summation in Equation~\eqref{eq:composite_image} results in equal contributions to the red, green, and blue channels:
\begin{equation}
\begin{split}
    \left(R[i,j], G[i,j], B[i,j]\right)  &=\frac{\sum_{k=0}^{N_{\text{sub}}-1} a[i,j] \mathbf{c}_k}{N_{\text{sub}}}  \\
    &=\frac{a[i,j] r}{N_{\text{sub}}} [1, 1, 1].
\end{split}
\label{eq:achromatic_condition}
\end{equation}
This equality produces a neutral gray or white appearance in the composite image, with saturation approaching zero. The preservation of this achromatic property in clean regions serves as a critical validation of the method's accuracy, confirming that the colorization process is selective and does not introduce spurious chromatic artifacts in the absence of interference.

\subsection{Parameter Selections and Practical Considerations}

The performance of the color mapping is influenced by several key parameters, which must be carefully selected to balance resolution, contrast, and computational efficiency. The number of subbands $N_{\text{sub}}$ directly affects the frequency resolution of the visualization; a larger $N_{\text{sub}}$ provides finer spectral discrimination but may lead to overly dense color gradients that reduce perceptual clarity. Empirical evaluation suggests that values within the range $N_{\text{sub}} \in [7, 11]$ offer an reasonable trade-off for Sentinel-1 SAR data. Finally, the percentile threshold $p$ must be chosen to effectively suppress noise without compromising the detection of weak interference signals. A value of 95\% has been found to provide a robust balance in typical scenarios.

\begin{figure}
  \centering
  \subfigure[\texttt{SIDS/20161119\_IW3\_B6789}]{
  \includegraphics[width=4.4cm,height=9.cm]{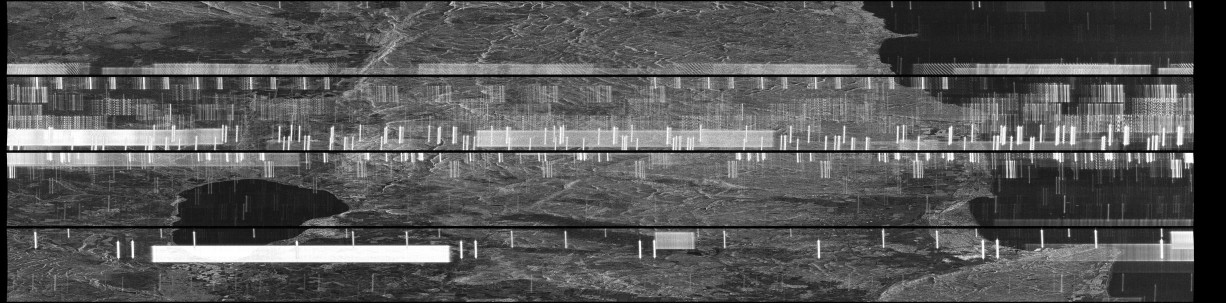}
  \includegraphics[width=4.4cm,height=9.cm]{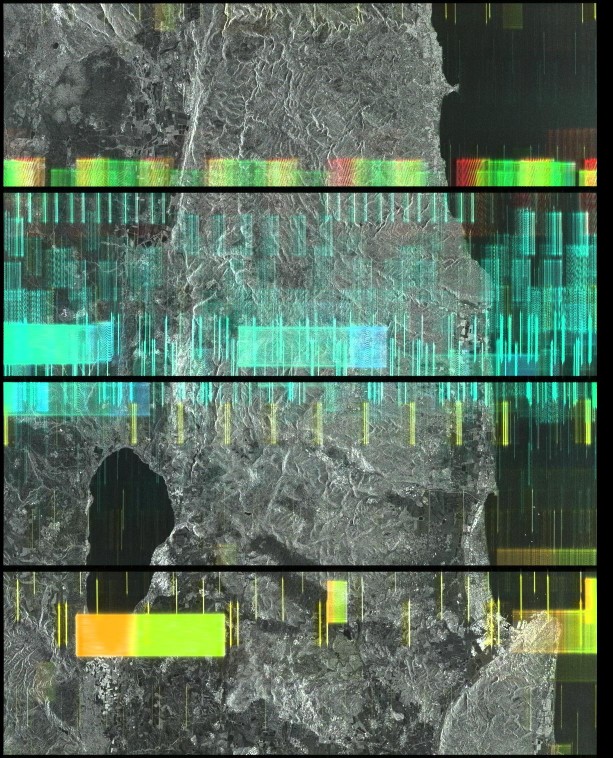}}
  
  \subfigure[\texttt{SIDS/20191031\_IW1}]{
  \includegraphics[width=4.4cm,height=2.25cm]{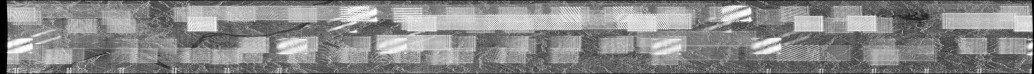}
  \includegraphics[width=4.4cm,height=2.25cm]{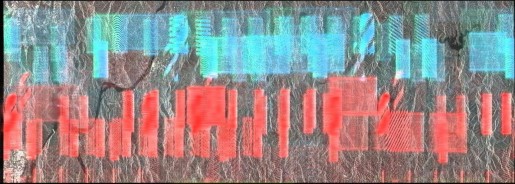}}
  
  \subfigure[\texttt{SIDS/20180731\_IW3\_B4}]{
  \includegraphics[width=4.4cm,height=2.25cm]{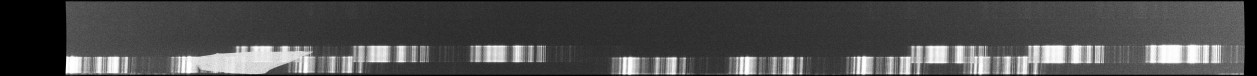}
  \includegraphics[width=4.4cm,height=2.25cm]{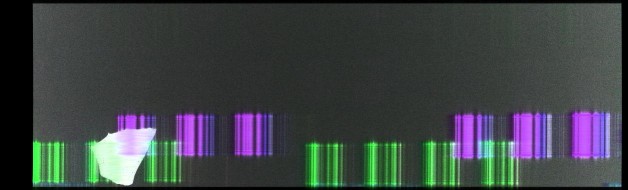}}
  
  \subfigure[\texttt{SIDS/20190411\_IW1\_B6}]{
  \includegraphics[width=4.4cm,height=2.25cm]{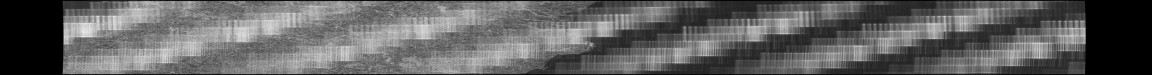}
  \includegraphics[width=4.4cm,height=2.25cm]{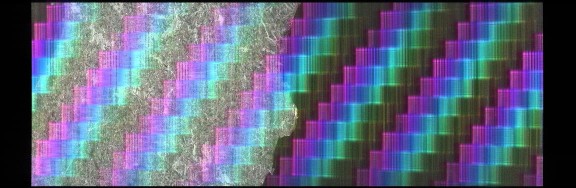}}
  
   \subfigure[\texttt{SIDS/20201110\_IW2\_B8}]{
   \includegraphics[width=4.4cm,height=2.25cm]{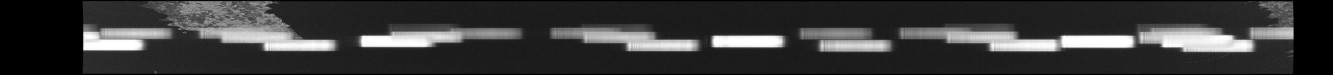}
  \includegraphics[width=4.4cm,height=2.25cm]{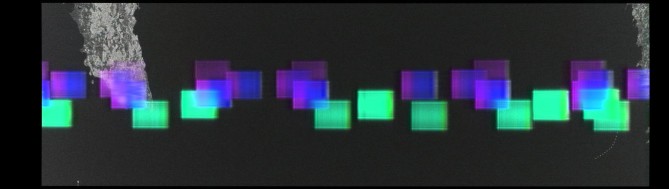}}
  \caption{Results (part 1) of spatial-spectral chromatic coding method on SIDS dataset. Left column: the gray-scale amplitude image; Right column: the proposed chromatic coding results.}\label{SIDS:1}
\end{figure}

\begin{figure}
  \centering
  \subfigure[\texttt{SIDS/20200801\_IW1\_B4}]{
  \includegraphics[width=4.4cm,height=2.25cm]{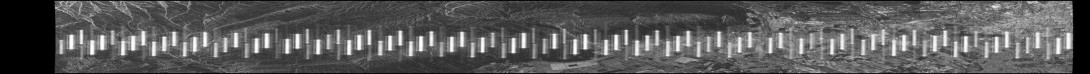}
  \includegraphics[width=4.4cm,height=2.25cm]{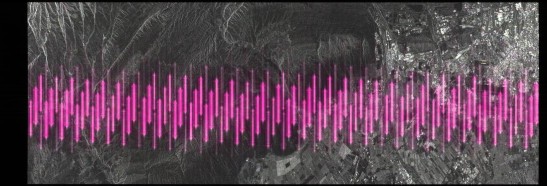}}
  
  \subfigure[\texttt{SIDS/20201212\_IW2\_B5}]{
  \includegraphics[width=4.4cm,height=2.25cm]{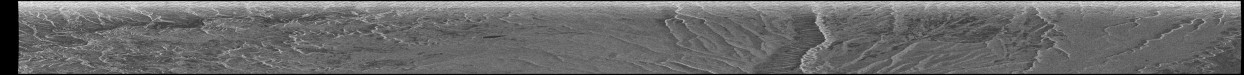}
  \includegraphics[width=4.4cm,height=2.25cm]{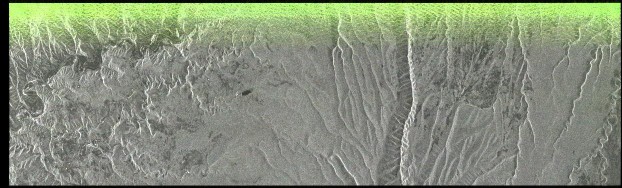}}
  
  \subfigure[\texttt{SIDS/20201212\_IW3\_B7}]{
  \includegraphics[width=4.4cm,height=2.25cm]{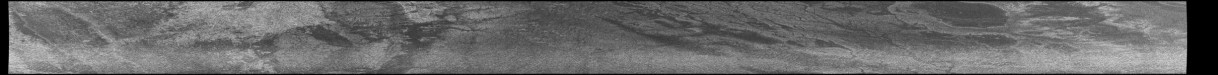}
  \includegraphics[width=4.4cm,height=2.25cm]{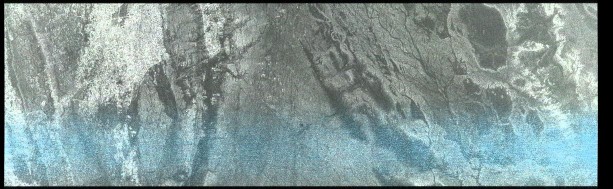}}
  
  \subfigure[\texttt{SIDS/20250525\_IW1\_B2}]{
  \includegraphics[width=4.4cm,height=2.25cm]{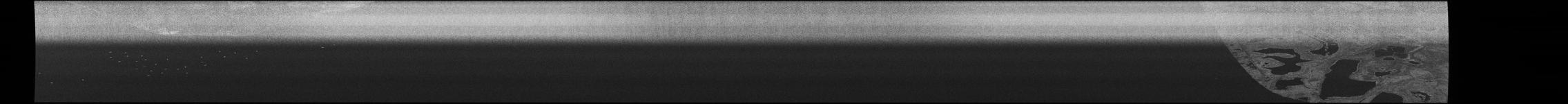}
  \includegraphics[width=4.4cm,height=2.25cm]{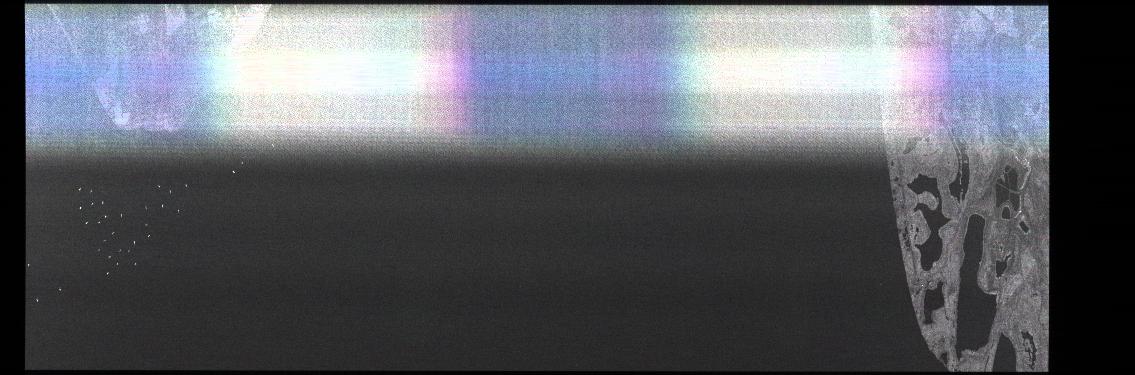}}
  
  \subfigure[\texttt{SIDS/20180812\_IW1\_B1}]{
  \includegraphics[width=4.4cm,height=2.25cm]{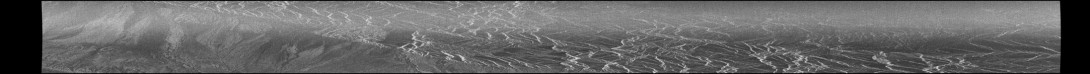}
  \includegraphics[width=4.4cm,height=2.25cm]{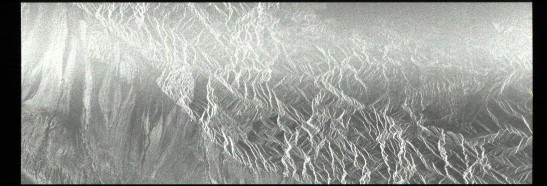}}
  
  \caption{Results (part 2) of spatial-spectral chromatic coding method on SIDS dataset. Left column: the gray-scale amplitude image; Right column: the proposed chromatic coding results.}\label{SIDS:2}
\end{figure}

\begin{figure}[]
  \centering
  \subfigure[\texttt{SIDS/20220310T153621\_IW2\_B89}]{
  \includegraphics[width=4.4cm,height=4.5cm]{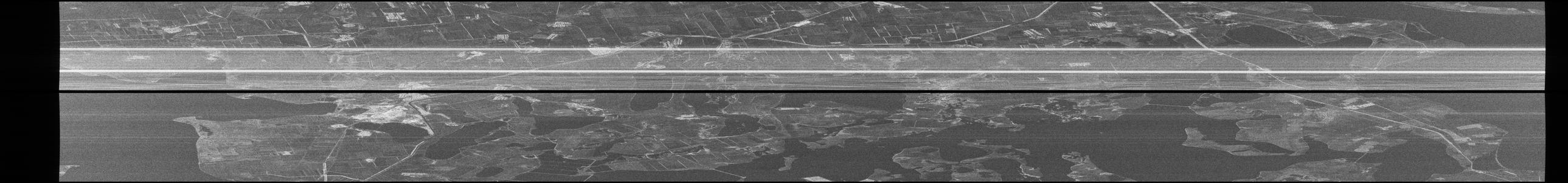}
  \includegraphics[width=4.4cm,height=4.5cm]{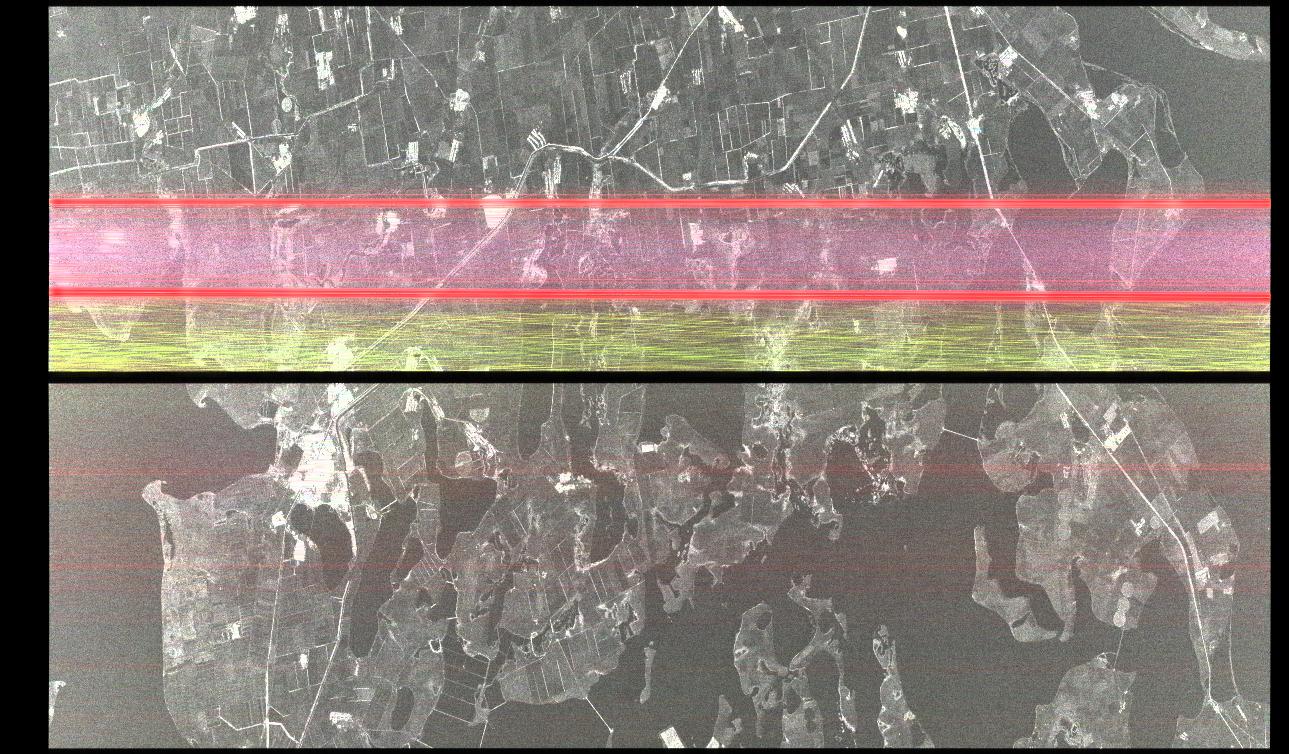}}
  
  \subfigure[\texttt{SIDS/20180402\_IW3\_B7}]{
  \includegraphics[width=4.4cm,height=2.25cm]{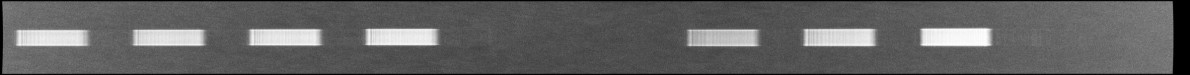}
  \includegraphics[width=4.4cm,height=2.25cm]{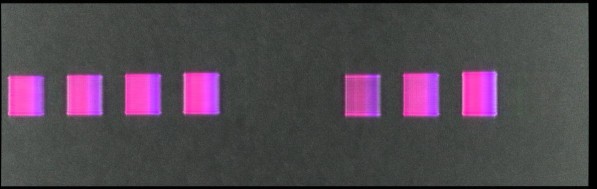}}
  
  \subfigure[\texttt{SIDS/20200420\_IW3\_B7}]{
  \includegraphics[width=4.4cm,height=2.25cm]{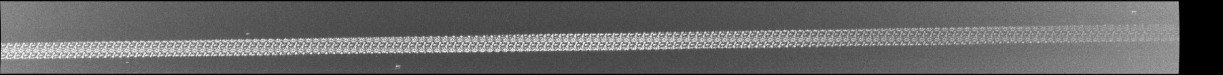}
  \includegraphics[width=4.4cm,height=2.25cm]{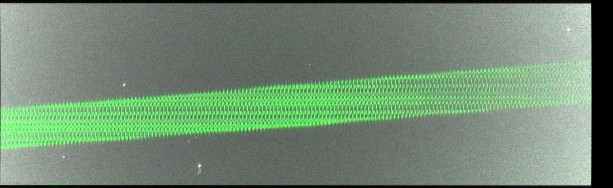}}
  
  \subfigure[\texttt{SIDS/20220310T153621\_IW1\_B9}]{
  \includegraphics[width=4.4cm,height=2.25cm]{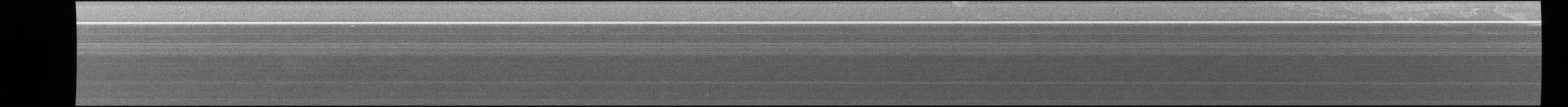}
  \includegraphics[width=4.4cm,height=2.25cm]{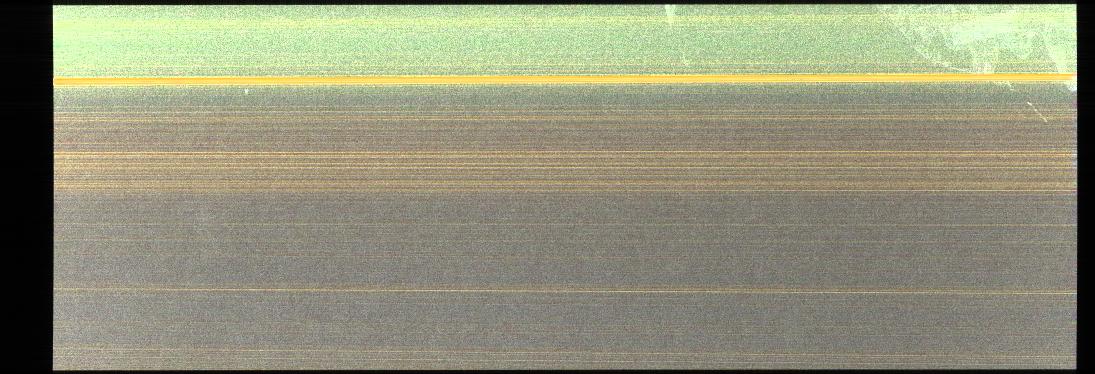}}
  
  \subfigure[\texttt{SIDS/20230925T153612\_IW1\_B6}]{
  \includegraphics[width=4.4cm,height=2.25cm]{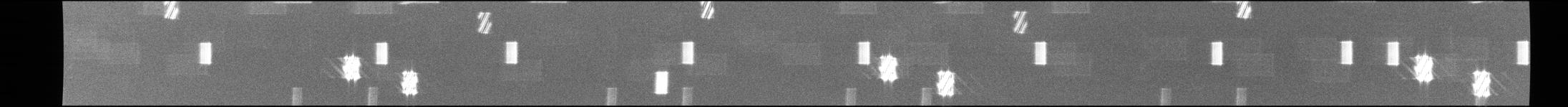}
  \includegraphics[width=4.4cm,height=2.25cm]{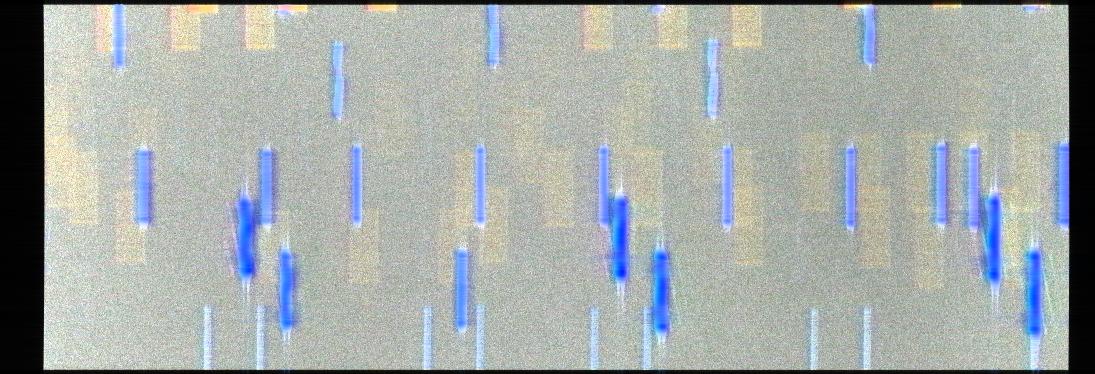}}
  
  \caption{Results (part 2) of spatial-spectral chromatic coding method on SIDS dataset. Left column: the gray-scale amplitude image; Right column: the proposed chromatic coding results.}\label{SIDS:3}
\end{figure}

\begin{figure*}[]
  \centering
  \subfigure[Blurring]{
  \includegraphics[width=8.8cm,height=3.cm]{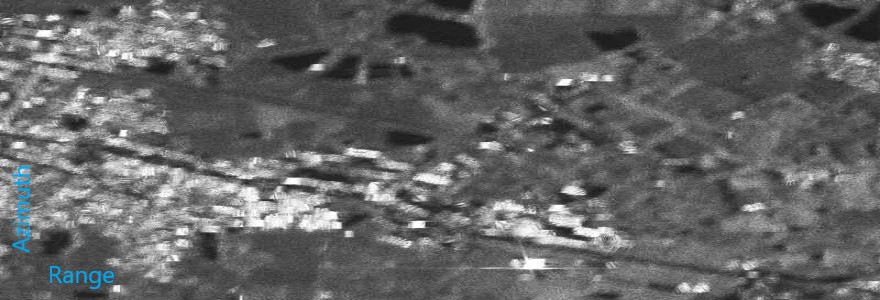}
  \includegraphics[width=8.8cm,height=3.cm]{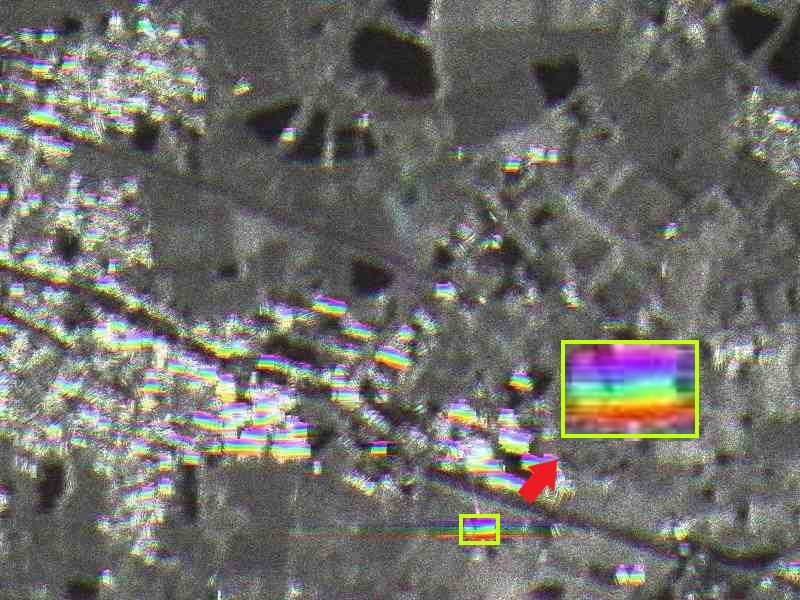}}
  
  \subfigure[Azimuth ambiguities]{
  \includegraphics[width=8.8cm,height=3.cm]{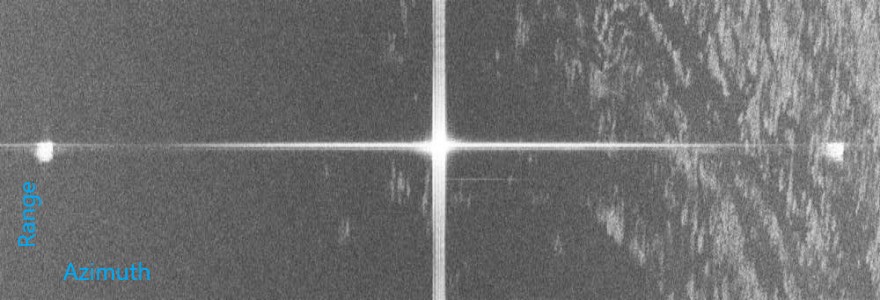}
  \includegraphics[width=8.8cm,height=3.cm]{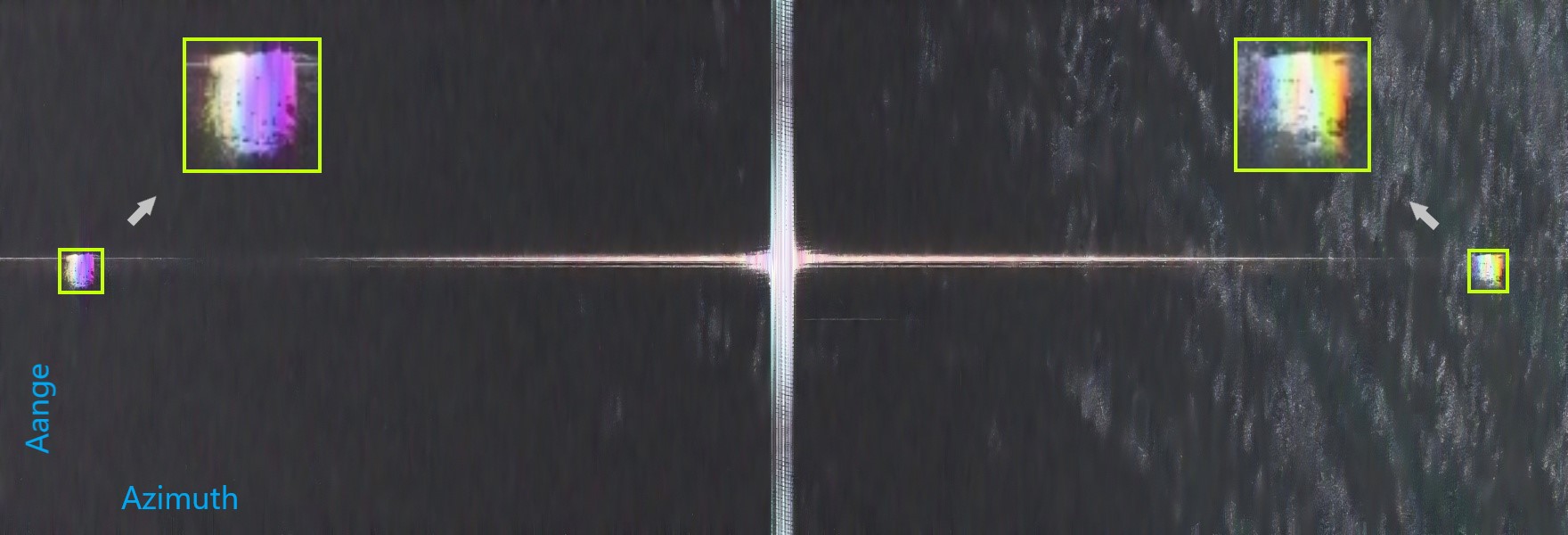}}
  
  \subfigure[Fast moving targets]{
  \includegraphics[width=8.8cm,height=3.cm]{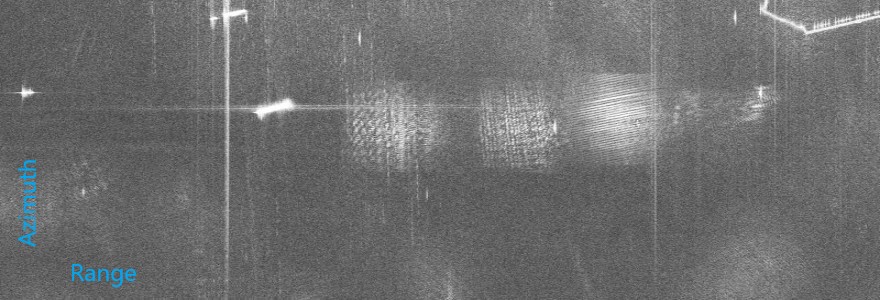}
  \includegraphics[width=8.8cm,height=3.cm]{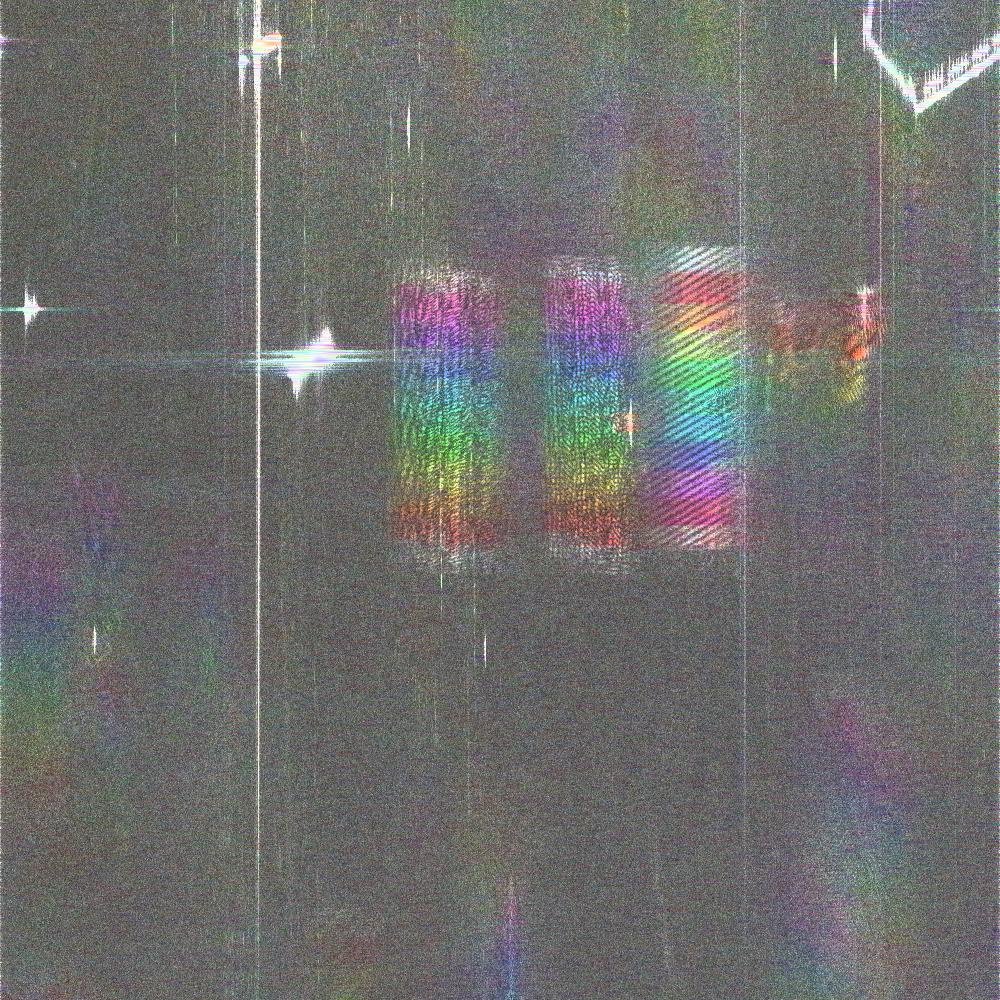}}

      \subfigure[L-band interference and fast moving targets]{
  \includegraphics[width=8.8cm,height=3.cm]{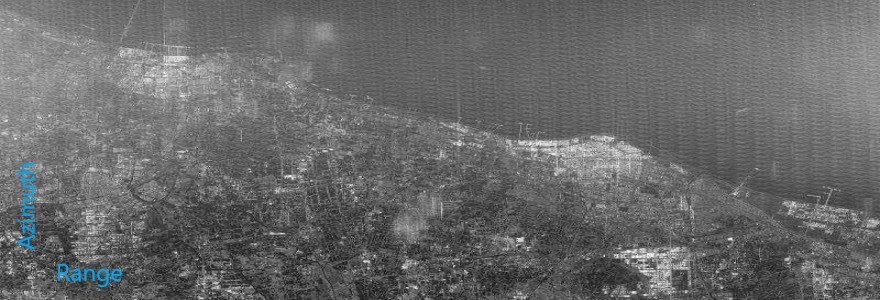}
  \includegraphics[width=8.8cm,height=3.cm]{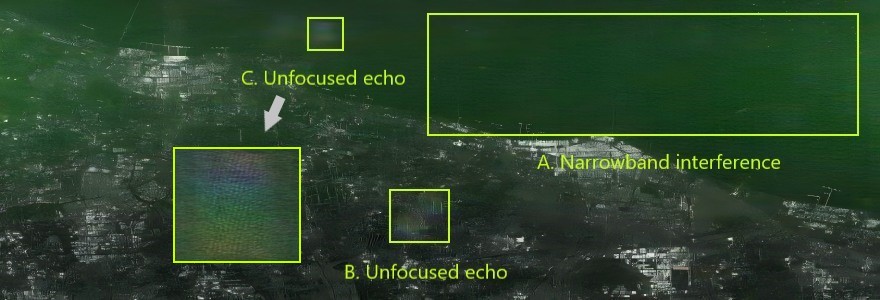}}
  
  \caption{Results of spatial-spectral chromatic coding method for the case of (a) blurring due to inaccurate Doppler centroid estimation, (b) azimuth ambiguities, (c) fast moving targets, and (d) L-band interference and fast moving targets.  Left column: the gray-scale amplitude image; Right column: the proposed chromatic coding results.}\label{fig:disortion}
\end{figure*}

\section{Experimental Results and Validation}
This section presents comprehensive experimental validation of the proposed chromatic visualization framework for SAR data analysis. We demonstrate the method's effectiveness in two critical application scenarios: complex interference characterization and unfocused signals (such as blurring and ambiguities). The experiments all use real SAR datasets. Through systematic evaluation, we show how the chromatic coding approach enables rapid identification and classification of interference patterns and distortions that remain indistinguishable in conventional grayscale imagery. 

\subsection{Dataset and Experimental Setup}
The validation of the proposed spatial-spectral decomposition framework was conducted using the SAR Interference Detection and Suppression (SIDS) dataset, a comprehensive benchmark developed by Nanjing University of Science and Technology. This dataset comprises 56 single-look complex (SLC) image slices from Sentinel-1 C-band SAR acquisitions, covering diverse coastal and terrestrial regions. The data exhibits a wide spectrum of interference characteristics, including pulsed/continuous waveforms, narrowband/wideband bandwidths, varying signal-to-interference ratios (SIR), and modulation styles ranging from noise to linear frequency modulation (LFM). Notably, the dataset includes both VH and VV polarizations, with VH-polarized data being predominant due to its heightened sensitivity to interference artifacts.

To ensure robust evaluation, we selected 10 representative scenes from the subset, which contains complex interference patterns with large dynamic ranges and diverse compositions. The experiments employed $N_{\rm sub} = 9$ spectral subband setting with $\xi = 1.16 ,1.34, 1.52$ (for IW1, IW2, IW3, respectively) oversampling rates,  look number $=[4, 20]$, and percentile parameter $p=95\%$.

\subsection{Qualitative Validation of Interference Visualization}

The results are provided in Figs. \ref{SIDS:1}, \ref{SIDS:2}, and \ref{SIDS:3}. It can be found that our method successfully transformed interference artifacts into diagnostically meaningful color patterns across diverse scenarios. In data {20161119\_IW3\_B6789} containing multiple narrowband and wideband interference, the method rendered interference as single color or continuously varying (due to LFM interference)  chromatic regions, clearly delineating them from natural features. In particular, the continuously varying colors of LFM interference regions clearly show the spectral sweep characteristic of the interference. For noise-like wideband interference (e.g., data  {20201212\_IW3\_B7}), the algorithm generated gradient cyan patterns that intuitively reflected the interference spectral sweep characteristics. 

A critical advantage emerged in scenes with coexisting interference types. For instance, data 20190607\_IW1\_B456789 contained both narrowband (red) and wideband  components, which the framework resolved as spatially distinct color zones. This capability addresses a key limitation of conventional amplitude-based visualization, where such overlapping signatures typically appear as monochromatic artefacts.

A notable case arises from full-band flat-spectrum noise interference, which exhibits uniform power distribution across the entire frequency spectrum. In particular, the SAR data 20180812\_IW1\_B1 contains partial contamination by this type of interference. The visualization presented in Fig. \ref{SIDS:2}(e) reveals that the affected regions appear as low-saturation white areas. This observation exactly aligns with our expectation, since such type of interference are called white noise.

\begin{table}[htbp]
\centering
\caption{SSIM between interference regions and clean regions for colored and grayscale images}
\label{tab:ssim_interference_clean}
\begin{tabular}{lcccccccccccccc}
\toprule
\textbf{Image Pair Id} & \textbf{1} & \textbf{2} & \textbf{3} & \textbf{4} & \textbf{5} & \textbf{6} & \textbf{7} \\
\midrule
Color image & 0.07 & 0.07 & 0.10 & 0.09 & 0.15 & 0.23 & 0.32  \\
Grayscale image & 0.21 & 0.20 & 0.28 & 0.19 & 0.53 & 0.48 & 0.45 \\
\bottomrule
\toprule
\textbf{Image Pair Id} &\textbf{8} & \textbf{9} & \textbf{10} & \textbf{11} & \textbf{12} & \textbf{13} & \textbf{14} \\
\midrule
Color image &  0.26 & 0.11 & 0.36 & -0.24 & 0.06 & 0.11 & 0.26 \\
Grayscale image  & 0.29 & 0.25 & 0.53 & 0.18 & 0.16 & 0.27 & 0.35 \\
\bottomrule
\end{tabular}
\end{table}

As shown in Table~\ref{tab:ssim_interference_clean}, the SSIM values differ between color and grayscale images when comparing interference and clean regions. The quantitative results suggest that the proposed colored visualization leads to improved visual identification of  interference regions from clean ones.

\subsection{Visualization of Unfocused Echoes: Blurring, Ambiguities, and Moving Targets Responses}
Fig.~\ref{fig:disortion} extends the visualization on unfocused signals, demonstrating the method's versatility in visualizing artifacts arising from unfocused echoes. The detailed result and analysis are listed below: 
\begin{itemize}
 \item Blurring due to inaccurate focusing parameters such as  Doppler centroid (Fig.~\ref{fig:disortion}a):
The grayscale image shows smeared structures with indistinct edges. Our chromatic coding result reveals cyan-green hues in blurred regions, indicating non-uniform spectral energy distribution caused by incorrect Doppler parameter estimation.

 \item Azimuth ambiguities (Fig.~\ref{fig:disortion}b):
The ambiguous targets appear as faint replicas in the grayscale image.
Our chromatic coding assigns distinct hues (e.g., red, green, blue) to ambiguous regions, reflecting their shifted spectral signatures due to Doppler aliasing. This enables differentiation from true targets. It is shown that azimuth ambiguities appear as symmetric color patterns (mirrored green-red pairs) displaced from true target positions, providing immediate visual cues about their ghost nature. 

 \item Unknown unfocused artefacts due to facts such as Doppler aliasing of moving targets   (Fig.~\ref{fig:disortion}c):
The unfocused energy manifests as smeared streaks in grayscale.
Our method renders these streaks in high-saturation colors (e.g.,  red to purple), indicating energy concentration in specific subbands caused by target motion-induced Doppler shifts.

 \item L-band interference with unknown unfocused artefacts (Fig.~\ref{fig:disortion}d):
This data has  L-band interference as well as unknown unfocused artefacts due to facts such as Doppler aliasing of moving targets. By our method, the interference regions appear green, while moving targets energy exhibit vivid colors (e.g., red, green), enabling simultaneous visualization of both phenomena.
\end{itemize}

It is notable that both the blurring, ambiguities, and unknown artefacts are unfocused signals. All of these artefacts cause localized color smearing (e.g., red to purple gradients) along the azimuth direction.

Across all cases, clean regions remain achromatic, validating the sum-white constraint. The results confirm that the method facilitate interference identification and analysis, providing a unified framework for visualizing and interpreting diverse interference (including unfocused signals.) in SAR imagery, and therefore offering radar analysts a powerful tool for SAR data analysis, diagnosis, and assessment without specialized processing.

In summary, the above visualization results of various interference and unfocused artefact patterns provides practical validation of the our model and method for enhanced interference visualization in SAR imagery.

\section{Conclusion}
This paper presents a spatial-spectral chromatic coding framework that transforms interference signatures in SAR single-look complex (SLC) images into perceptually distinct color features. By establishing a rigorous mathematical linkage between interference spectral characteristics and color channels, the method fundamentally redefines interference from an obscuring artifact into an information-rich diagnostic tool. The core innovation lies in decomposing an SLC image into multiple spatial-spectral sub images through subband-level Fourier analysis, then mapping these sub images to the RGB color space via a set of basis color matrices uniformly sampled in the HSV space. This approach preserves the full complex data integrity while enabling intuitive visualization of interference patterns that were previously indistinguishable in conventional amplitude displays. Experimental validation on the real SAR datasets demonstrates the framework’s effectiveness in enhancing interference visualization. The color-encoded representations consistently reveal spatial distribution patterns and spectral characteristics of interference sources, providing analysts with actionable insights for signal origin identification and mitigation strategy development.

We believe that the proposed visualization paradigm establishes a new standard for SAR data interpretation, transforming complex interference phenomena into accessible visual diagnostics. By integrating spectral decomposition with standard color palette, this work bridges the gap between radar image measurements and human perceptual understanding, offering a foundational tool for advancing interference analysis in radar imaging systems.

\bibliographystyle{IEEEtran}
\bibliography{refs,refs2}

\end{document}